\documentclass[11pt]{article}
\usepackage{moriond,epsfig}
\usepackage{graphicx}
\usepackage{epsf}
\usepackage{epsfig}
\usepackage{color}
\usepackage{bm}

\def\bea{\begin{eqnarray}}
\def\eea{\end{eqnarray}}

\let \to=\rightarrow

\newcommand{\EIPWA}{MIPWA}

\newcommand{\SA}{\ensuremath{s_{K\pi}}}
\newcommand{\SAA}{\ensuremath{s}}
\newcommand{\SAB}{\ensuremath{s^{\prime}}}
\newcommand{\SAk}{\ensuremath{s_{K\pi}^k}}

\newcommand{\swave}{\ensuremath{ S\hbox{-wave}}}
\newcommand{\pwave}{\ensuremath{ P\hbox{-wave}}}
\newcommand{\dwave}{\ensuremath{ D\hbox{-wave}}}
\newcommand{\sdash}{\ensuremath{ S\hbox{-}}}
\newcommand{\pdash}{\ensuremath{ P\hbox{-}}}
\newcommand{\ddash}{\ensuremath{ D\hbox{-}}}

\newcommand{\KBs}{\ensuremath{K_0^*(1430)}}
\newcommand{\KAp}{\ensuremath{K^*(892)}}
\newcommand{\KBp}{\ensuremath{K_1^*(1410)}}
\newcommand{\KCp}{\ensuremath{K_1^*(1680)}}
\newcommand{\KAd}{\ensuremath{K_2^*(1430)}}

\newcommand{\sst}{\scriptscriptstyle}

\newcommand{\half}{\ensuremath{{1\over 2}}}

\newcommand{\MeVcc}{\ensuremath{\hbox{MeV}/c^2}}
\newcommand{\GeVcc}{\ensuremath{\hbox{GeV}/c^2}}

\let\mathrm=\rm

\newcommand{\pip}{\pi^+}

\newcommand{\Km}{K^-}

\newcommand{\Dp}{D^+}

\begin{document}
\vspace*{4cm}
\title{\swave\ $\Km\pip$ SYSTEM IN $\Dp\to\Km\pip\pip$ DECAYS
FROM FERMILAB E791}

\author{B. T. MEADOWS}

\address{Department of Physics, University of Cincinnati, \\
Cincinnati, OH, 45221-0011, USA \\[6pt]
\centerline{\rm Representing the Fermilab E791 Collaboration}}

\maketitle\abstracts{
A new approach to the analysis of three body decays is presented.
Model-independent results are obtained for the \swave\ $K\pi$ amplitude 
as a function of $K\pi$ invariant mass.  These are compared with results 
from $\Km\pip$ elastic scattering, and the prediction of the Watson 
theorem, that the phase behavour be the same below $K\eta^{\prime}$ 
threshold, is tested.  Contributions from $I=\half$ and $I={3\over 2}$ 
are not resolved in this study.  If $I=\half$ dominates, however, the 
Watson theorem does not describe these data well.}

\section{Introduction}
Decays of heavy-quark mesons are regarded as a potential source of 
information on the light-quark mesons they produce.  For decays
to three pseudo-scalar final states, kinematics and angular momentum 
conservation favor production of \swave\ systems, so improving our
knowledge of the particularly confusing scalar meson ($J^P=0^+$)
spectrum may be possible when the anticipated large, clean samples
of such decays of $D$ mesons from the $B$ factories and the Tevatron
collider become available.
Extracting this information has, however, been done in model-dependent
ways that make assumptions about the scalar states observed.  Such
assumptions can influence the results, so new approaches are required.

In this paper, we present a model-independent approach to the \swave\
system in a study of the decays $\Dp\to\Km\pip\pip$
\footnote{Throughout the paper, charge conjugate states are implied 
unless explicitly stated otherwise.}
observed in data from Fermilab experiment E791.
%in a way that can provide model-independent information on the
%$\Km\pip$ \swave\ system.
We compare the \swave\ amplitudes
so obtained with our earlier, isobar model analysis
\cite{Aitala:2002kr} and also with $\Km\pip$ scattering.

Measurements of $\Km\pip$ scattering come principally
from SLAC experiment E135 (LASS)
\cite{Aston:1987ir},
and cover the invariant mass range only above
825~\MeVcc.  Data exist below this range, but with less
precision
\cite{Estabrooks:1977xe}.
More information in the low mass region is required if the
possibility of the existence of a $\kappa$ state is to be
properly evaluated.

\section{Data Sample}
\begin{figure}[hbt]
 \centerline{%
 \epsfig{file=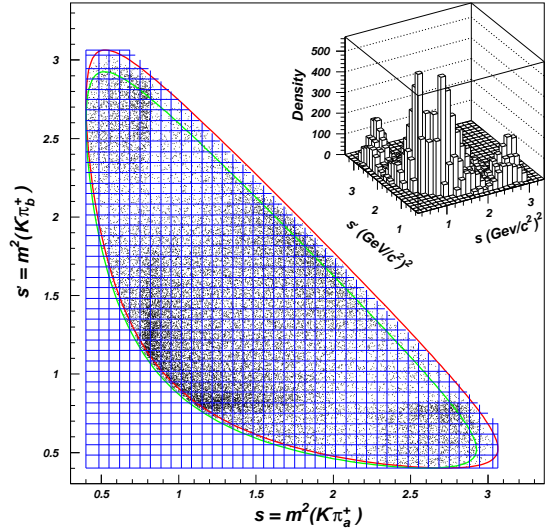,width=0.5\textwidth,angle=0}}
 \caption{Dalitz plot for $\Dp \to \Km\pip_a\pip_b$ decays.  The
  squared invariant mass $\SAA$ is plotted against
  $\SAB$.  The plot is symmetrized, each event
  appearing twice.  Lines in both directions indicate values equally
  spaced in squared effective mass at each of which the \swave\
  amplitude is determined by the model-independent partial wave
  analysis (\EIPWA) described in section
  \ref{sec:method}.  Kinematic boundaries for the Dalitz plot are
  drawn for three-body mass values $M=1.810$ and $M=1.890$~\GeVcc,
  between which data are selected for the fits.
  \label{fig:dalitz_plot}}
\end{figure}

The selection process for events used in this analysis is described
in Ref.~\cite{Aitala:2002kr}.  A signal consisting of 15,079 
$\Dp\to\Km\pip_a\pip_b$ decays, with a purity of $\sim 94$\%, is
obtained.  Fig.~\ref{fig:dalitz_plot} shows the Dalitz plot with
$\Km\pip$ squared invariant mass \SAA\ plotted vs. \SAB.
%%%%%%%%%%%%%
Horizontal (and the symmetrized vertical) bands corresponding
to the \KAp\ resonance are clearly seen.  A striking and
complex pattern of both constructive and destructive interference
is seen near 2 (\GeVcc)$^2$ due to either \KBs, 
\KBp\, or \KAd.
%A further enhancement appears in the top left (and bottom right)
%corners of the plot.
There is also evidence for \KCp, difficult to see due to smearing 
of the Dalitz plot boundary resulting from
the finite resolution in the three-body $\Dp$ mass.

The most striking effect observed is the asymmetry in the \KAp\
bands, most easily described by interference with a significant
\swave\ contribution to the decay.  In this paper we are able to
extract information on the \swave\ using the \KAp, and also the
other well established resonances in the Dalitz plot, as an
interferometer.

\section{Method}\label{sec:method}
In Ref.~\cite{Aitala:2002kr}, as in most earlier analyses of
$D$ decays to three pseudo-scalar particles $ijk$, we use the
``isobar model".  Details of this are given in
\cite{Aitala:2000xu}.
In this model, the decay amplitude ${\cal A}$ is described by
a sum of quasi two-body terms $D\to R+k,~R\to i+j$, in each of
the three channels $k=1,2,3$:
 \bea\label{eq:isobarmodel}
   {\cal A} &=& d_0 e^{i\delta_0} + 
        \sum_{n=1}^N d_{n} e^{i\delta_{n}}
        ~{F_{\sst R}(p,r_{\sst R},J)\over
        m_{\sst R_n}^2-s_{ij}-im_{\sst R_n}\Gamma_{\sst R_n}(s_{ij})}
        \times
        F_{\sst D}(q,r_{\sst D},J)
        ~M_J(p,q)
 \eea
In this, $s_{ij}$ is the squared invariant mass of the $ij$ system.
$J$ is the spin, $m_{\sst R_n}$ the mass and $\Gamma_{\sst R_n}(s_{ij})$
the width of each of the $N$ resonances $R_n$ seen to be
contributing to the decay.  $F_{\sst R}$ and $F_{\sst D}$
are form factors, with effective radius parameters
$r_{\sst R}$ and $r_{\sst D}$, for all $R_n$ and for the parent $D$
meson, respectively.  $p$ and $q$ are momenta of $i$ and $k$,
respectively, in the $ij$ rest frame.  $M_J(p,q)$ is a factor
introduced to describe spin conservation in the decay.  The complex
coefficients $d_n e^{i\delta_n}~(n=0,N)$ are determined by the $D$
decay dynamics and are parameters estimated by a fit to the data.  The first,
non-resonant ($NR$) term describes direct decay to ${i+j+k}$ with no
intermediate resonance, and $d_0$ and $\delta_0$ are assumed to be
independent of $s_{ij}$.
For $\Dp\to\Km\pip_a\pip_b$ decays we Bose-symmetrize
${\cal A}$ with respect to interchange of $\pip_a$ and $\pip_b$.

In Ref.~\cite{Aitala:2002kr} we reported that the $NR$ term was 
smaller than previously thought, and that a further term,
parametrized as a new $J=0$ resonance $\kappa(800)$ with
$m_{\sst R}=(797\pm 19\pm 43)$~\MeVcc and
$\Gamma_{\sst R}=(410\pm 43\pm 87)$~\MeVcc, gave a much better
description of the data.  Here, we examine the $\Km\pip$ \swave\
in a model-independent way.  The \swave\ part of
Eq.~\ref{eq:isobarmodel} (all terms with $J=0$, including the
$NR$ term) is factored
\bea\label{eq:swave}
   {\cal S} &=& \hbox{S}(\SA)
        \times
        M_0^{\sst R}(p,q)
        F_{\sst D}(q,r_{\sst D})
        ~=\hbox{Interp} \left(c_k e^{i\gamma_k}\right)
        \times
        M_0^{\sst R}(p,q)
        F_{\sst D}(q,r_{\sst D})
\eea
into a partial wave S$(\SA)$, describing $\Km\pip$ scattering,
and the product $M_0^{\sst R}(p,q)F_{\sst D}(q,r_{\sst D})$
describing the $D$ decay.  S$(\SA)$ is interpolated between a set
of points $c_k e^{i\gamma_k}$ defined at 40 $\Km\pip$ invariant
mass squared values $\SAk$ indicated by the lines in
Fig.~\ref{fig:dalitz_plot}.  Each $c_k$ and $\gamma_k$ is regarded
as an independent parameter determined by the data.

We factor the \pdash\ and \ddash\ reference waves in the same way:
\bea\label{eq:refwaves}
   {\cal P} &=& \hbox{P}(\SA)
        \times
        M_1^{\sst R}(p,q)
        F_{\sst D}(q,r_{\sst D})
   ~;~
   {\cal D} ~=~ \hbox{D}(\SA)
        \times
        M_2^{\sst R}(p,q)
        F_{\sst D}(q,r_{\sst D}),
\eea
however, we parametrize the partial waves P$(\SA)$ and D$(\SA)$
exactly as in Eq.~\ref{eq:isobarmodel}.

%We find the $(c_k,\gamma_k)$ values that provide
%the best description of the Dalitz plot density
%$|{\cal S} + ({\cal P}+{\cal D})|^2$.
We make an unbinned likelihood fit to the data, Using the method
described in
Ref.~\cite{Aitala:2002kr}.
This incorporates an incoherent background function describing
the 6\% of our sample not corresponding to true $D$ decays.
We measure 86 parameters - all $(c_k,\gamma_k)$ and the coefficients
$d_k e^{i\delta_k}$ for \KAp, \KCp\ in the \pwave\ and \KAd\
in the \dwave.  For the \KAp, we define $d_k e^{i\delta_k}=1$ 
to provide the reference phase.

The fit results in an excellent description of the data.
Comparison of the observed and predicted population of the
Dalitz plot gives a $\chi^2$ probability of 50\% for 363 bins.

\section{Results}
\begin{figure}[hbt]
 \centerline{%
 \epsfig{file=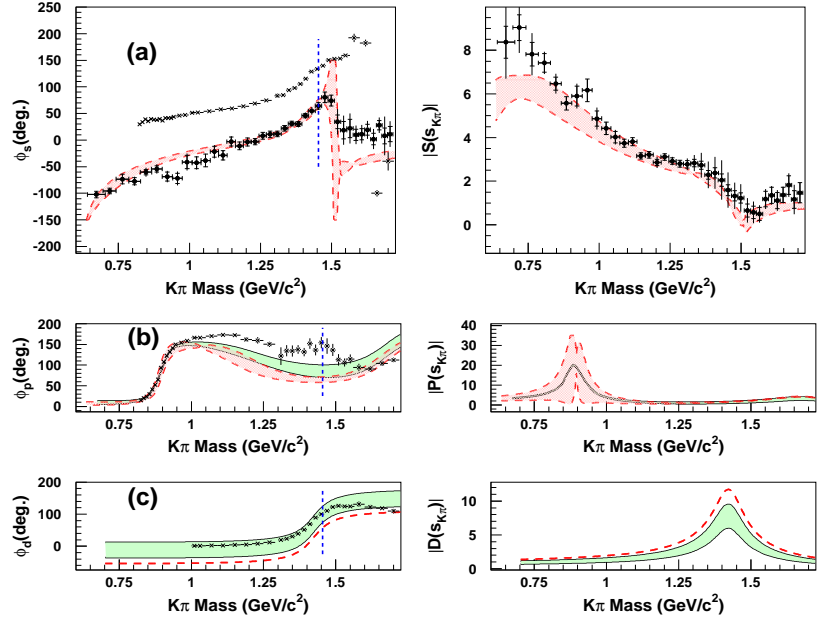,
         width=0.5\textheight,
         height=0.39\textheight,angle=0}}
 \caption{%
 \label{fig:solution0}
  (a) Phases (left) and magnitudes (right) of \swave\ amplitudes
  for the $\Km\pip$ systems from $\Dp\to\Km\pip\pip$ decays obtained
  from the \EIPWA\ fit described in the text.
  The effect of adding systematic uncertainties in quadrature with
  statistical errors is indicated by extensions on the error bars.
  The \pwave\ and \dwave\ amplitudes are plotted in (b) and (c),
  respectively, as curves
  derived from the isobar parameters used in these waves, and from
  the full error matrix resulting from the fit.  Shaded areas
  represent one standard deviation limits on the amplitudes.  In
  all plots, the dashed curves show one standard deviation
  limits for the prediction of the isobar model fit.
  $I=1/2$ phase measurements from the LASS experiment, with
  vertical and horizontal error bars, are included in the phase
  plots.  The vertical lines mark the $\Km\eta^{\prime}$
  threshold, the upper limit of elastic scattering.
% and the range where the Watson theorem applies.
% Arrows indicate the range of measurements for which the Watson
% theorem predicts that the data should lie on this curve.
}%
\end{figure}

The \sdash, \pdash\ and \dwave s resulting from the fit are shown
in Fig.~\ref{fig:solution0}.  They are compared with the
model-dependent fit from Ref.~
\cite{Aitala:2002kr}.
The main \swave\ features of both fits agree well.  Resonant
fractions and the total \swave\ fraction (about 75\%) also
agree within statistical limits.

We turn now to a comparison of the \swave\ amplitudes S$(\SA)$
measured here with the amplitudes $T(\SA)$ measured in $\Km\pip$
elastic scattering.
% by the LASS collaboration \cite{Aston:1987ir}.
We expect, for each partial wave $J$ (for each iso-spin $I$)
that 
$\hbox{S}(\SA)={\sqrt{\SA}/ p^{(J+1)}}Q(\SA)\hbox{T}(\SA)$
where $Q(s_{K\pi})$ describes the dependence of $\Km\pip$
production in $D$ decays on \SA.  The Watson theorem
\cite{Watson:1952ji}
requires that, provided there is no re-scattering of the
$\Km\pip_a$ from $\pip_b$, that $Q$ is a real function, so
that phases found in $D$ decay should match those in $\Km\pip$
elastic scattering data.

$I=1/2$ phases measured by LASS are plotted in
Fig.~\ref{fig:solution0}.  There is a large offset in the
\swave, about 75$^{\circ}$, not seen in \pdash\ or \dwave s.
The shapes of \sdash\ and \pwave s are also not the same.
Unless significant admixture of $I=3/2$ $\Km\pip$ production
occurs, these results suggest that the conditions for the Watson
theorem are not met in these data.

\section*{Acknowledgments}
We thank members of the LASS collaboration for making 
their data available to us.
We gratefully acknowledge the assistance of the staffs of Fermilab
and of all the participating institutions.  This research was
supported by the Brazilian Conselho Nacional de 
Desenvolvimento Cient\'{\i}fico e Tecnol\'{o}gico,
CONACyT (Mexico), FAPEMIG
(Brazil), the Israeli Academy of Sciences and Humanities,
%PEDECIBA (Uruguay),
the U.S. Department of Energy, the U.S.-Israel
Binational Science Foundation, and the U.S. National Science 
Foundation.  Fermilab is operated by the Universities Research
Association for the U.S. Department of Energy.

\section*{References}
\bibliography{paper}

\end{document}